\documentclass[aps,pra,reprint,groupedaddress,showpacs]{revtex4-1}%
\usepackage{amsmath}
\usepackage{graphicx}
\usepackage{textcomp}
\usepackage{hyperref}
\usepackage{amsfonts}
\usepackage{amssymb}%
\providecommand{\U}[1]{\protect\rule{.1in}{.1in}}
\begin{document}
\title{Reexamination of energy flow velocities of non-diffracting localized waves}
\author{Peeter Saari$^{1,2}$, Ott Rebane$^{1,3}$, and Ioannis Besieris$^{4}$}
\email{Corresponding author: peeter.saari@ut.ee}
\affiliation{$^{1}$Institute of Physics, University of Tartu, W. Ostwaldi 1, 50411, Tartu, Estonia}
\affiliation{$^{2}$Estonian Academy of Sciences, Kohtu 6, 10130 Tallinn, Estonia}
\affiliation{$^{3}$LDI Innovation Ltd, Osmussaare 8, 13811 Tallinn, Estonia}
\affiliation{$^{4}$The Bradley Department of Electrical and Computer Engineering, Virginia
Polytechnic Institute and State University, Blacksburg, Virginia 24060, USA}
\date{\today}

\begin{abstract}
A universal relation has been established between the local energy transport
velocity along the direction of propagation and the group velocity of scalar
and vector-valued propagation-invariant spatiotemporally localized
superluminal and subluminal electromagnetic waves in free space. Under
specific restrictions, this relationship is very closely valid for physically
realizable almost propagation-invariant spatiotemporally confined subluminal
and superluminal electromagnetic fields. In both cases, although the group
velocity may be either superluminal or subluminal, the universal relation is
in accord with the well-established result that the upper limit of the energy
transport velocity is $c$, the speed of light in vacuum.

\end{abstract}

\maketitle

\section{Introduction}

The propagation speed of pulses of structured light has attracted much
attention in recent years
\cite{Gio2015,Horv2015,Bareza,Bouch,MinuComm1,Alf,MinuComm2,Faccio2,MinuPRA2018,KondakciArbitV,AbourVisC2019}%
. It is well known that while different velocities associated with light
propagation are equal to the universal constant $c$ in the case of
one-dimensional plane waves in vacuum, it is not so when the plane waves are
propagating in dispersive media, where the group velocity can take any value
below or above $c$. In particular, relations between the group velocity,
energy transport velocity and the pulse's time of flight become complicated
and even controversial, see, e.g., Refs. \cite{7kiirust,Peatross2000} and
review \cite{MilonniReview}.

In the case of 2- or 3-dimensional structured light pulses, even if they
propagate in empty space, certain space-time couplings can emulate temporal dispersive properties. Such couplings materialize through correlation of
the spatial frequencies involved in the construction of the structured pulses
with the temporal frequencies $\omega$ constituting the pulse temporal
profile. If for all Fourier constituents of the pulse the correlation consists
in a linear functional dependence between $\omega$ and the component $k_{z}$ of
the wave vector, which lies in the direction of propagation of the pulse, then
the pulse is called propagation-invariant. This means that its intensity
profile, or spatial distribution of its energy density, does not change in the
course of propagation---it does not spread either in the lateral or in the
longitudinal direction (or temporally). In reality such a non-diffracting
non-spreading propagation occurs over a large but still finite distance,
because the functional dependence is not strict for practically realizable
(finite-energy and finite aperture) pulses.

The first versions of such propagation-invariant localized pulsed waves were
theoretically discovered in the late 1980-ies and since then a massive literature
has been devoted to them, see collective monographs \cite{LWI,LW2} and reviews
\cite{DonelliSirged,revPIER,revSalo,MeieLorTr,KiselevYlevde,AbourClassif2019}.
The realizability of them in optics was first demonstrated in
Ref.~\cite{PRLmeie} for the example of so-called Bessel-X pulse which is the
only propagation-invariant pulsed version of the monochromatic Bessel beam
introduced in Ref.~ \cite{Durnin}. The group velocity of Bessel-X pulses
exceeds $c$, i.e., it is superluminal in empty space without\ the presence of
any resonance medium. This strange property has been widely discussed in the
literature referred to above and was experimentally verified by several groups
\cite{exp2,exp3,meieXfemto,meieOPNis} for cylindrically symmetric 3D pulses
and recently for 2D (light sheet) counterparts of such superluminal pulses
\cite{AbourClassif2019,Xsheet}.

Motivated by the growing interest in studying the propagation and applications
of structured light pulses in general, and by recently introduced techniques
of generation of pulsed light sheets with space-time couplings in particular,
the following question arises. How is the group velocity of
propagation-invariant pulses related---and whether it is related at all---to
the energy flow velocity in them? Definitely the statement \textquotedblleft
if an energy density is associated with the magnitude of the wave ... the
transport of energy occurs with the group velocity, since that is the rate of
which the pulse travels along\textquotedblright\ (citation from
Ref.~\cite{Jackson}, section 7.8) cannot hold if the group velocity exceeds
$c$. Indeed, very general proofs show that no electromagnetic field can
transport energy faster than $c$ \cite{Lekner2002,Yannis2008,YannisLWII}, i.e., even in the case of superluminal pulses. On the other hand, how should one
comprehend the situation where energy flows slowly, thus as if lagging behind
the pulse?

There are few calculations of the Poynting vector and energy density of
electromagnetic propagation-invariant pulses
\cite{Recami1998,Faccio2010,Salem2011}. To our best knowledge, there is only
one work where the energy flow velocity of a vectorially treated superluminal
Bessel-X light pulse has been calculated \cite{Mugnai2005}. In this work the
velocity is found to be equal to $c$ from the symmetry ($z$) axis up to near the first zero of the Bessel function. As we will see below, this result is
not exact and was obviously obtained due to carrying out the final evaluation
numerically in a paraxial geometry. The authors of Ref.~\cite{Mugnai2005}
conclude that "it is not clear what kind of physical mechanism makes the energy velocity different from the phase and group ones".

We have shown earlier \cite{OttMag,PIERS2013} that the spatial distribution of
the Poynting vector, the energy density and its transport velocity, calculated
numerically by means of scalar approximation formulas for the Bessel beams,
practically coincide with the results of an exact vectorial approach, and the
value of the velocity is slightly below $c$ for various propagation-invariant
scalar fields.

The main objective of the present study is to evaluate analytically how the
energy flow velocity is related---if it is related at all---to the group
velocity in the case of various propagation-invariant vectorial and scalar
fields. Let us note that in this paper we deal primarily with instantaneous
energy flow velocity which, as a matter of fact, does not depend on time in a
frame copropagating with the propagation-invariant field. Since commonly the
average energy flow velocity per period of a time-harmonic EM field is
considered in the literature, for which the name "energy transport velocity"
is used, following Ref.~\cite{Kaiser2011} we will avoid this term if
time-harmonic fields are not considered. We will frequently use simply the
short form "energy velocity".

The paper has been organized as follows. In Section II we reproduce the proof
that the upper limit for the energy flow velocity in any EM field is $c.$ The
same for any scalar field is presented in the Appendix. In Section III we derive
a rather universal relation between the group velocity and the energy velocity
of propagation-invariant transverse magnetic (TM) 2D and 3D superluminal fields.
We start with the 2D case, i.e., with light sheets not only for reasons of
simplicity and transparency but also having in mind that pulsed light sheets
have also practical value, e.g., in microscopy, and are presently studied
intensively
\cite{KondakciArbitV,AbourVisC2019,AbourClassif2019,Xsheet,KondakciSSelfH}.
Section IV deals with energy velocities of several known cylindrical scalar and
vectorial superluminal fields. Section V is devoted to subluminal pulsed
fields and, in particular, to a propagation-\textit{variant }so-called pulsed
Bessel beam, which is generated by a diffractive axicon and is essential for
applications. In Section VI we discuss the interpretation and nature of the
obtained universal expression for the axial energy flow velocity in terms of
the theory of special relativity and in terms of the normalized impedance of
non-null EM fields. Finally, we speculate on the reasons why the value of the energy velocity is different from that of the group velocity.

\section{Upper limit of energy flow velocity}

The local energy flow velocity, as it is well known, is given by ratio of
energy flux (Poynting vector) and the electromagnetic energy density as (SI
units)%
\begin{equation}
\mathbf{V}=2c^{2}\frac{\mathbf{E}\times\mathbf{B}}{\mathbf{E}^{2}%
+c^{2}\mathbf{B}^{2}}~. \label{Vgeneral}%
\end{equation}
The magnitude of this quantity cannot exceed the universal constant $c$, the speed of light in
vacuum. Indeed, by using the general vector identity%

\[
\left(  \mathbf{E}\times\mathbf{B}\right)  ^{2}=\mathbf{E}^{2}\mathbf{B}%
^{2}-\left(  \mathbf{E}\cdot\mathbf{B}\right)  ^{2}%
\]
one can write \cite{Lekner2002,Yannis2008,YannisLWII,Kaiser2011}%
\begin{equation}
1-\frac{\mathbf{V}^{2}}{c^{2}}=\frac{\left(  \mathbf{E}^{2}-c^{2}%
\mathbf{B}^{2}\right)  ^{2}+4c^{2}\left(  \mathbf{E}\cdot\mathbf{B}\right)
^{2}}{\left(  \mathbf{E}^{2}+c^{2}\mathbf{B}^{2}\right)  ^{2}}~. \label{v<c}%
\end{equation}
The right-hand side of Eq.~(\ref{v<c}) is nonnegative and as a consequence
$\left\vert \mathbf{V}\right\vert \leq c$. Luminal velocity $\left\vert
\mathbf{V}\right\vert =c$ is applicable only to TEM waves; also, to
\textit{null} EM fields \cite{BB2003,YannisLWII} a trivial example of which is
a single plane wave. Since the two terms in the numerator on the right-hand
side of Eq.~(\ref{v<c}) are known to be Lorentz invariant, the property of
subluminality or luminality of $\mathbf{V}$ does not depend on the speed of a
reference frame.

Optical fields, especially paraxial ones, can be in good approximation
described by a single scalar function $\psi(x,y,z,t)$. In this case, the
Poynting vector $\mathbf{S}$ and energy density $w$ are given by the
expressions \cite{MW}
\begin{subequations}
\label{scalarSw}%
\begin{align}
\mathbf{S}  &  \mathbf{=-}\alpha\left(  \dot{\psi}^{\ast}\mathbf{\nabla}%
\psi+\dot{\psi}\mathbf{\nabla}\psi^{\ast}\right)  ~,\\
w  &  =\alpha\left(  c^{-2}\dot{\psi}\dot{\psi}^{\ast}+\mathbf{\nabla}%
\psi\cdot\mathbf{\nabla}\psi^{\ast}\right)  ~,
\end{align}
where the dot denotes derivative with respect to time, and the asterisk
complex conjugation, $\mathbf{\nabla}$ is the gradient operator, and $\alpha$ is a
positive constant whose value depends on the choice of units. The local energy
flow velocity is again given by the ratio $\mathbf{V}=\mathbf{S}/w,$ the
constant $\alpha$ cancels out and, as proved in the Appendix, the limit
$\left\vert \mathbf{V}\right\vert \leq c$ holds for scalar fields as well.

\section{Relation between group velocity and energy velocity of
propagation-invariant electromagnetic fields.}

For a field to be propagation-invariant, it must depend on the propagation
distance $z$ and time through the difference $z-vt$, where $v$ is the group
velocity. Consider, for example, the scalar wave known under name "fundamental
X-wave" first derived in \cite{Lu-X,Zio1993}:%
\end{subequations}
\begin{equation}
\psi_{x}\left(  \rho,z,t\right)  =\frac{1}{\sqrt{\rho^{2}+\left[
a+i\tilde{\gamma}\left(  z-vt\right)  \right]  ^{2}}}\label{Xpsi}%
\end{equation}
in polar coordinates. Here, $v$ is a superluminal speed of\ propagation of the
whole pulse and $\tilde{\gamma}\equiv1/\sqrt{(v/c)^{2}-1}$ is the superluminal
version of the Lorentz factor, $c$ being the speed of light in vacuum. The
positive free parameter $a$ determines the width of the unipolar
Lorentzian-like temporal profile of the pulse on the $z$ axis. The
double-conical spatial profile of the field looks like the letter "X" in a
meridional plane $(\rho,z)$. In order to get a dc-free optical wave containing
a number of cycles, one has to take temporal derivatives of correspondingly
high order from Eq.~(\ref{Xpsi}). If such field is expanded into monochromatic
plane waves or Bessel beams, $v$ turns out to be also the phase velocity
(along the $z$ axis) of all monochromatic constituents of the field, and is
given as $v=c/\cos\theta$, where $\theta$ is the common inclination angle of
all the constituents with respect to the axis $z$. The angle $\theta$ is
called the Axicon angle in the case of a Bessel-beam expansion. As $\cos
\theta\leq1$, such a field can propagate only with a superluminal group
velocity. In this Section we will derive a universal relation between the
superluminal group velocity and subluminal energy velocity of
propagation-invariant electromagnetic fields.

\subsection{The case of 2D fields (light sheets)}

We start with fields that do not depend on one lateral, say $y,$ coordinate.
Although such 2D fields are simpler, they possess the same properties as 3D ones.

Consider a TEM electromagnetic (generally pulsed) 2D wave propagating along
the positive $z$ direction in vacuum,%

\begin{equation}
\mathbf{E}(x,z,t)=U(z-ct)\mathbf{e}_{x},~\mathbf{B}(x,z,t)=c^{-1}%
U(z-ct)\mathbf{e}_{y},\label{EjaB}%
\end{equation}
where SI units are assumed, with $\varepsilon_{0}\mu_{0}=1/c^{2},$ $U$ is an
arbitrary real localized function of \textit{one} argument, and $\mathbf{e}%
_{x}$ and $\mathbf{e}_{y}$ are unit vectors of a right-handed rectangular
coordinate system. The energy flux density and the energy density of the field
are given by%
\begin{align}
\mathbf{S} &  =c^{2}\varepsilon_{0}\ \mathbf{E\times B}=c\varepsilon_{0}%
U^{2}(z-ct)\mathbf{e}_{z}\ ,\label{Poynting}\\
w &  =\frac{1}{2}\varepsilon_{0}\mathbf{E}^{2}+\frac{1}{2}\varepsilon_{0}%
c^{2}\mathbf{B}^{2}=\varepsilon_{0}U^{2}(z-ct)\ .\label{Energia}%
\end{align}
Hence, the energy velocity is $\mathbf{V}=\mathbf{S}/w=c\mathbf{e}_{z}$, which
is a well-known result.

Let us take now a symmetrical pair of plane waves---the propagating direction
of the first one lies on the $(x,z)$ plane and is inclined by angle $+\theta$
with respect to the $z$-axis, and the second one by angle $-\theta$ on the
same plane. In this case, the coordinate $z$ in Eq.~(\ref{EjaB}) is replaced
by $z\cos\theta+/-x\sin\theta$ for a member of the pair, respectively. The
components of vectors $\mathbf{E}$ and $\mathbf{B}$ for both waves transform
also according to the rules of rotation around the axis $y$. For the
polarizations given in Eq.~(\ref{EjaB}), the magnetic field remains polarized
along the $y$ axis. However, the electric field has both $x$ and $z$
components. Thus, we are dealing with a TM electromagnetic field. The
resulting expressions for $\mathbf{S}$ and $w$ are rather cumbersome;
therefore, we give them here only for the case $x=0$:%
\begin{align}
\mathbf{S} &  =4c\varepsilon_{0}\cos\theta\ U^{2}(z\cos\theta-ct)\mathbf{e}%
_{z}\ ,\label{Steljel}\\
w &  =\varepsilon_{0}\left[  \cos2\theta+3\right]  U^{2}(z\cos\theta
-ct)\ .\label{wteljel}%
\end{align}
{\small  }At any spatiotemporal point with a fixed value of \ $x$, the
quantities $\mathbf{S}$ and $w$ and, hence, the vector field of energy velocity
depend on \textit{the propagation distance and time} solely through the
difference $z\cos\theta-ct$ in the argument of the function $U$. Thus the
vectors fields of energy flux and energy velocity and the scalar field of energy density all move without any change in the $z$-direction with velocity
$v=c/\cos\theta$.

From Eqs.~(\ref{Steljel}-\ref{wteljel}), with the help of some trigonometry we
get for the quantity of our primary interest---the energy velocity on the
propagation axis $z$ (shortly: the axial velocity)---the following expression:%
\begin{gather}
\mathbf{V}=\mathbf{S}/w=V_{z}\ \mathbf{e}_{z}\ ,\qquad V_{z}\equiv
V=cR(\theta),\nonumber\\
R(\theta)\equiv\frac{2\cos\theta}{\cos^{2}\theta+1}\ .\label{R()}%
\end{gather}
We see that the energy velocity does not depend on the function $U$, has only
the axial component on the propagation axis and takes \textit{only subluminal
value}\emph{s} in the interval from $c$ to $0$ depending on the angle $\theta$
as depicted in Fig.~1. 
\begin{figure}
\centering
\includegraphics[width=8.5cm]{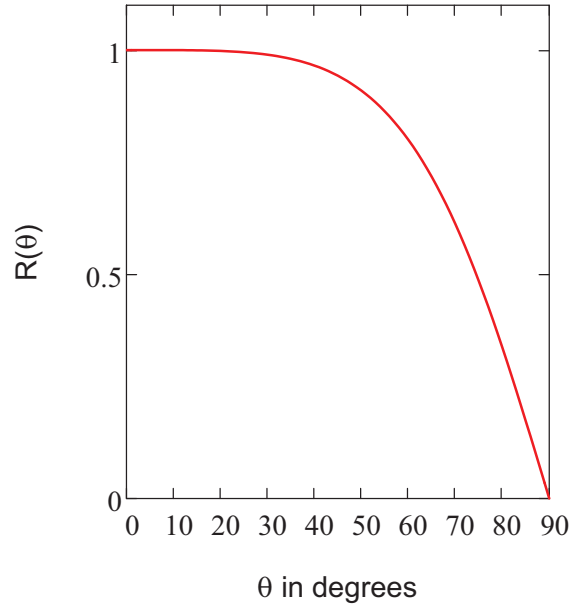}%
\caption{The subluminality factor in Eq.~(\ref{R()})
representing the energy flow velocity (in units of $c$) along the propagation
axis $z$ in dependence on the inclination angle $\theta$ of the plane waves.}
\end{figure}
The case $\theta=90^{\circ}$ corresponds to a standing
wave where, as it is well known, energy does not flow. More precisely: in this
case the plane $(y,z)$ is a node plane outside of which the energy flows back
and forth along the $x$ axis in accordance with the time function $U^{2}.$
(For harmonic time dependence, the behavior of the energy flow instantaneous
velocity of standing waves has been thoroughly studied, e.g., in
\cite{Kaiser2011}). We will see in the following that the obtained ratio
$R(\theta)$ of the energy axial velocity to the universal constant $c$ is not
peculiar to the given simple model 2D field but holds generally for
propagation-invariant fields. Moreover, if we introduce the normalized
velocity $\beta\equiv v/c$, we can rewrite Eq.~(\ref{R()}) in following two
forms:%
\begin{equation}
R(\beta)\equiv\frac{2\beta^{-1}}{\beta^{-2}+1}=\frac{2\beta}{\beta^{2}%
+1}~.\label{R(b)}%
\end{equation}
The last equality is remarkable and we shall comment on it in the discussion
provided in Sec. VI.

In order to get an idea about the energy velocity vectors outside of the axial
region, the velocity field is depicted in Fig.~2 for the case of the pulse
wavefunction $U$ comprised of a single positive half-period of a cosine.
Outside the crossing region the energy flows perpendicularly to the pulse
front with velocity $c$ as expected. In the central region, the pulses sum up
resulting in almost a 4-fold (if the angle $\theta$ is small) increase of both
the energy flux density and the energy density, while the ratio of these two
quantities---the energy velocity---is smaler than $c$. Since the horizontal
axis of Fig.~2 represents the propagation variable $\zeta\equiv z\cos
\theta-ct$, the plots can be interpreted either as "snapshots in flight" made
at a fixed value of time $t$, or as plots on the $(-t,x)$-plane made at a fixed
value of the coordinate $z$. From the latter interpretation it follows
immediately that for all spatial points, except for those on the $z$ axis, the
$x$-component of the Poynting vector as well as $V_{x}$ reverse their sign
with increasing of time. In contrast, $S_{z}$ and $V_{z}$ remain non-negative
for all values of $x$, $z$, and $t$.

\begin{figure}
\centering
\includegraphics[width=9cm]{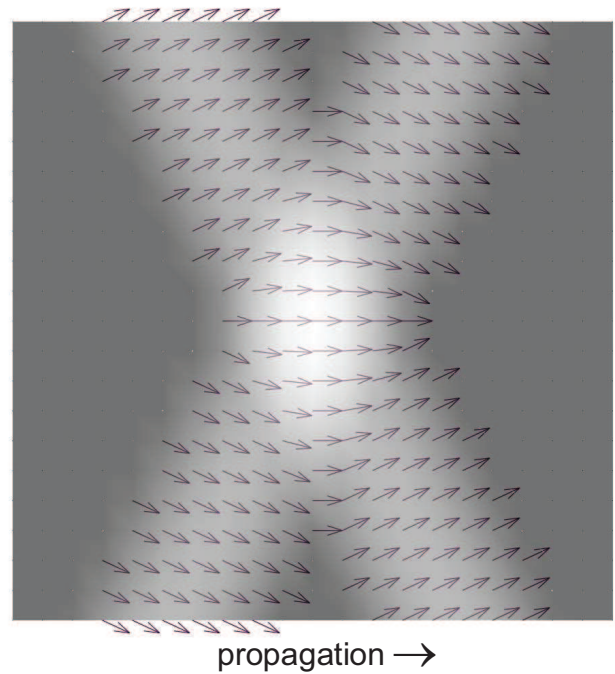}%
\caption{The field of energy flow velocities formed by
the crossing of two unipolar half-cycle pulses, depicted by arrows on a grid
of $21\times21$ points. The length of arrows represents the magnitude of the
vectors, which equals to $c$ on the branches of the X-profile and is smaller
than $c$ on the $z$ axis according to Eq.~(\ref{R()}). The $x$ axis is
vertical. The horizontal axis represents the propagation variable $z\cos
\theta-ct$. For $\theta=25^{\circ}$, the energy velocity on the axis $z$ is
$V_{z}\equiv V=0.99518$ in accordance with Eq.~(\ref{R()}). The vector field
plot is superimposed by a semitransparent greyscale surface (grid $21\times
21$) plot of the square root of the energy density distribution. The velocity
vectors are shown only for regions where the energy density exceeds 0.1\% of
its maximum value.}
\end{figure}

In Fig.~3, the velocity field is depicted for the case where the pulse
wavefunction $U$ comprises a cosine in the interval from $-3\pi/2$ to $3\pi
/2$. We see an embryonic pattern of interference between crossing harmonic
plane waves. While the velocity equals to $c$ and is directed perpendicularly
to the X-branches outside of the interference region, it is much smaller and
points in various directions in regions of destructive interference. Other
features are the same as in Fig. 2.
\begin{figure}
\centering
\includegraphics[width=9cm]{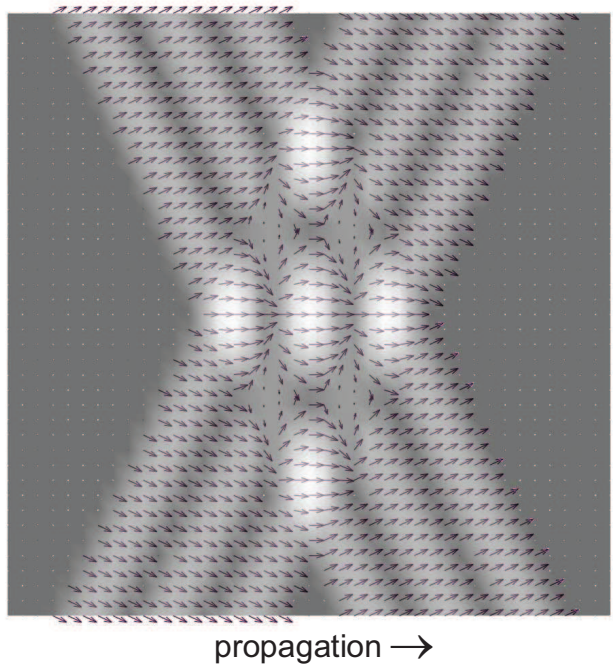}%
\caption{The field of energy flow velocities and the
square root of the energy density distribution formed by the crossing of two
bipolar 1.5-cycle pulses. The grid of $\ $the plots consists of $41\times41$
points. For other characteristics see caption of Fig.~2.}
\end{figure}

Despite the rather coarse grid, Fig. 3 indicates some subtleties in the
behavior of the energy velocity at locations of minima of the energy density
in the case when the function $U$ contains an oscillating factor. A detailed
numerical and analytical study with $U$ taken as a sine or cosine function
reveals the following: (i) on a line $\zeta=\zeta_{0}$ corresponding to a
zero of the sine or cosine, the Poynting vector vanishes identically for all
values of $x$ in the region of interference of the two waves, while the energy
density is nonzero, except for the value $x=0$; (ii) the local energy velocity
is zero in such planes, except at the points on the $z$ axis where
Eq.~(\ref{R()}) holds as it does generally on the $z$ axis; (iii) at the
points $(\zeta=\zeta_{0},x=0)$ the velocity makes a jump between zero and the
value given by Eq.~(\ref{R()}); (iv) which of the two values the velocity
takes depend on the order in taking the double limit $\zeta\rightarrow
\zeta_{0},x\rightarrow0$. Although such a discontinuity is unphysical, it is
not a serious problem, since the energy density vanishes at the discontinuity
points and the velocity is therefore undefined in a physical sense anyway
(mathematically there is the $0/0$-uncertainty).

\subsection{ Generalization to 3D cylindrical fields}

Although propagation-invariant light sheets have become a subject of intensive
study recently, as mentioned in the Introduction, 3D counterparts of them, in
particular cylindrically symmetric ones, have been of main interest. One of
the reasons is that the energy density in the central spot of a monochromatic
$J_{0}$-Bessel beam, as well as in the apex of the double-conical profile of a
X-type pulse, exceeds considerably (more than four times) the energy density
outside the center, which further decreases inversely proportionally with the
distance from the propagation axis.

Do the relations Eq.~(\ref{R()}) or Eq.~(\ref{R(b)}) also hold for 3D
propagation-invariant fields? The latter can be considered as summing up the
pairs of plane waves considered in the previous subsection whereas the axis
$x$ there takes all values of the azimuthal angle $\phi$ $\in\left[
0\ldots\pi\right]  $ around the axis $z$. Since the quantities $S$ and $w$ are
not linear with respect to fields $\mathbf{E}$ and $\mathbf{B}$, it is far
from being obvious that the expressions Eq.~(\ref{Steljel})-(\ref{wteljel})
hold also for such a cylindrically symmetric superposition of the fields. On
the $z$ axis the EM fields of a pair of plane waves considered above are given
by
\begin{subequations}
\begin{align*}
\mathbf{E}_{p}(x,z,t)|_{x=0} &  =2\cos\theta\,U(z\cos\theta-ct)\mathbf{e}%
_{x}\ ,\\
\mathbf{B}_{p}(x,z,t)|_{x=0} &  =2c^{-1}U(z\cos\theta-ct)\mathbf{e}_{y}\ .
\end{align*}
Let us now take another such pair for which the axis $x$ is rotated by an
angle $\phi$ around the axis $z$. If we denote the corresponding coordinate
transformation $3\times3$ matrix by $T(\phi)$, we can express the
$1/2$-weighted sum of the fields of the two pairs as $\frac{1}{2}\left[
1\mathbf{+}T(\phi)\right]  \mathbf{E}_{p}$ , $\frac{1}{2}\left[
1\mathbf{+}T(\phi)\right]  \mathbf{B}_{p}$ and calculate the flux and energy
density. The resulting expressions turn out to be the same as
Eqs.~(\ref{Steljel}) and (\ref{wteljel}) but both multiplied by a factor
$(\cos\phi+1)$ which cancels out from the expression of $V$. As in the case of
obtaining Eq.~(\ref{R()}), the square of the function $U$ also cancels out
from the ratio of the energy flux and density. Thus, Eq.~(\ref{R()}), with the
subluminality factor $R(\theta)$, remains valid for the resultant field of two
or more pairs of plane waves irrespective of the azimuthal angles between
their directions. Likewise, for a rotationally symmetric superposition of the
pairs, one has to integrate $T(\phi)\mathbf{E}_{p}$ and $T(\phi)\mathbf{B}%
_{p}$ over $\phi$ in the interval $\left[  0\ldots\pi\right]  $ and calculate
the flux and energy density in the resultant fields. The results coincide with
Eqs.~(\ref{Steljel}) and (\ref{wteljel}) multiplied by $4$. Hence, the factor
$R(\theta)$ given by Eq.~(\ref{R()}) expresses also in the case of
rotationally symmetric propagation-invariant 3D waves the dependence of the
energy flow velocity along the propagation axis on the Axicon angle $\theta.$

\section{Energy flow velocity of superluminal propagation-invariant
electromagnetic and scalar fields}

\subsection{Fields with fixed value of Axicon angle}

It is interesting first to verify whether the formula in Eq.~(\ref{R()}) or
Eq.~(\ref{R(b)}) holds in the case of the best known superluminal
fields---Bessel beams, Bessel-X and X-waves, where not only the intensity or
energy density but also the field itself is propagation-invariant. All these
comprise cylindrically symmetric superpositions of plane waves directed at a
fixed angle $\theta$ with respect to axis $z$ and differ only by the temporal
wave profile $U$. The function $U$ contains, respectively, infinitely many
(Bessel beam) or few cycles (Bessel-X) or a single unipolar Lorentzian-like pulse (X-wave)
and cancels out in the ratio of energy flux and density.
Therefore the energy velocity on the propagation axis (i.e., where $\rho
\equiv\sqrt{x^{2}+y^{2}}=0)$ of all these waves should be given by
Eq.~(\ref{R()}) or Eq.~(\ref{R(b)}).

Expressions for the time-averaged Poynting vector and energy density for
electromagnetic Bessel beams of the zeroth order can be found in
\cite{Ari1993}. Examination of the complicated Eq.~(44) (for $\mathbf{S}$) and
Eq.~(43) (for $w$) derived there for a plane-polarized $\mathbf{E}$, bearing
in mind that the ratio $\beta/k$ is equal to our $\cos\theta$, shows that on
the axis the ratio $\mathbf{S}/w$ indeed turns out to be same as our
Eq.~(\ref{R()}). Examination of Eq.~(51) (for $w$) and Eq.~(52) (for
$\mathbf{S}$) derived there for the case of a circularly polarized Bessel beam
results in the same conclusion. In the last case of polarization the authors
of Ref.~\cite{Ari1993} have found that at the radial distances from the $z$
axis which correspond to the zeros of the Bessel function, the $z$-component
of the Poynting vector assumes slightly negative values. Since it means also
negative values of $V_{z}$, for the sake of comparison we carried out
calculations of the time-averaged Poynting vector, energy density and velocity
of TM Bessel beams obtained with the Hertz vector, whose $z$-component is
given by a scalar Bessel beam field of arbitrary order $m$. Our results are
the following: (i) the $z$-component of the velocity does not assume negative
values at any radial distance $\rho$ from the $z$-axis; (ii) as $\rho$
increases the velocity oscillates (in accordance with the behavior of the
Bessel function) while the maximum values are given by Eq.~(\ref{R()}); (iii)
for $m>0$ the first maximum is at $\rho=0$, i.e., the formula Eq.~(\ref{R()}
holds on the $z$-axis, while for $m=0$ the velocity is zero on the $z$-axis.

The Poynting vector and energy density for electromagnetic fields derived from
a Hertz potential given by a scalar ultrabroadband X-shaped wave, first
derived in \cite{Lu-X,Zio1993}], were calculated in \cite{Recami1998}, see
Eq.~(6) therein. Again, the angular dependence of the ratio of the two
quantities coincides with our Eq.~(\ref{R()}). However, since the EM field
vectors are obtained from Hertz potentials through spatial and temporal
derivatives, the central maximum of the X-shaped Hertz potential turns into
zero values for some field components, as well as the Poynting vector, on the
$z$ axis. In contrast, the first-order scalar ultrabroadband X-wave, which is
azimuthally asymmetric (has a factor $\exp i\varphi$), is zero in the center.
In this case the axial component of the Poynting vector does not vanish in the
center of the pulse as one can see from Eq.~(14) of Ref.~\cite{Salem2011}
where the Poynting vector of such X-wave has been calculated for a general
(TM+TE) polarization.

In order to clarify when our formula in Eq.~(\ref{R()} or Eq.~(\ref{R(b)}
holds and when not, we calculated the energy velocity for different EM-fields
derived from X-shaped Hertz potentials. Without loss of generality, we
restricted ourselves to the TM field case. The results are the following.
\end{subequations}
\begin{enumerate}
\item In the case of the azimuthally symmetric scalar potential $\psi
_{x}\left(  \rho,z,t\right)  $ given in Eq.~(\ref{Xpsi}), the local energy
flow velocity turns out to be zero at the center of the pulse, but in the
central cross-sectional plane it increases with the distance from the $z$ axis
and approaches its maximum on a ring of radius $\rho=\sqrt{2}a$, where the
formula in Eq.~(\ref{R()}) holds. Such a behavior is similar to that of the
case of the Bessel beam of order $m=0$ described above.

\item In the case of the azimuthally asymmetric scalar potential given by
$\psi_{x}^{as}\left(  \rho,z,\varphi,t\right)  =\psi_{x}\left(  \rho
,z,t\right)  ^{3}\rho\exp(i\varphi)$, the formula in Eq.~(\ref{R()}) holds at
the center of the pulse, as well as on a ring of radius $\rho=2a$, where its
second maximum is located.

\item In the case of the azimuthally asymmetric scalar potential used to form
the Hertz vector in \cite{Salem2011}, the formula in Eq.~(\ref{R()}) holds at
the center of the pulse. The same holds for another azimuthally asymmetric
scalar potential taken for the Hertz vector in \cite{Recami1998}.
\end{enumerate}

\subsection{Fields with frequency-dependent Axicon angle}

Not all non-diffracting fields can be represented as angular superpositions of
plane wave pulses as shown earlier, or---in the case of spectral
representation of cylindrical fields---as superpositions of monochromatic
Bessel beams whose axial wavenumbers are proportional to the frequency. More
general non-diffracting fields, where not the field itself but only its
intensity (energy flux and/or density) is propagation-invariant, can be
represented as angular superpositions of \textit{tilted} pulses or---in
spectral terms---as superpositions of Bessel beams where the axial wavenumber
depends linearly but not simply proportionally on the frequency (see, e.g.,
\cite{LWI,meieLWIs,MeieLorTr}). This means that the angle $\theta$ is not
fixed any more and becomes a function of frequency within the spectral band of
the pulse.

Since (i) directed optical EM fields are in good approximation describable as
scalar fields and (ii) there are many studies of scalar non-diffracting fields
in the literature but few studies of their EM counterparts, in what follows we
deal primarily with the energy velocity of scalar fields. However, first we
must answer the question to what extent is the scalar treatment justified in
calculations of the energy velocity. It is easy to check that in the case of a
pair of scalar plane waves (light sheets), the same expressions for the axial
velocity given in Eqs. (\ref{R()}) and (\ref{R(b)}) follow from Eq.
(\ref{scalarSw}).

Since a single-frequency Bessel beam is the constituent of all cylindrical
non-diffracting pulses, we carried out numerically comparisons between energy
velocity fields of a scalar Bessel beam and vectorial (EM) Bessel beams
\cite{OttMag,PIERS2013} The main result is that the velocity field
calculated using the scalar approximation practically coincides with those
calculated for EM\ Bessel beams of different polarizations, except for small
off-axis regions around minima of energy density, where the discrepancy is
about a few percent of $c$ if $\theta<20^{\circ}$ and much less at paraxial
values of $\theta$.

Bessel beams considered so far have infinite aperture and therefore cannot be
generated in reality. To check that Eq.~(\ref{R()} works also with realistic
Bessel beams, we applied Eq.~(\ref{scalarSw}) to a so-called Bessel-Gauss beam
(see \cite{PorrBessGauss} and Refs. therein), which is a cylindrically
symmetric superposition of Gaussian beams propagating under the Axicon angle
$\theta$ with respect to the $z$ axis and have a superluminal group velocity
in the waist region. Our result is that Eq.~(\ref{R()}) holds if
$\theta<15^{\circ}$ which is understandable since the Bessel-Gauss beam is a
solution of the paraxial wave equation.

The first example of a superluminal scalar field with frequency-dependent
Axicon angle is the so-called Focused X Wave (FXW) \cite{revPIER},
possibilities of optical generation of which have been considered in
Ref.~\cite{meieFXW}. The expression for the FXW reads
\begin{align*}
\psi_{fxw}\left(  \rho,z,t\right)   &  =\psi_{x}\left(  \rho,z,t\right)
\times\\
&  \exp\left[  \frac{-|k|}{\psi_{x}\left(  \rho,z,t\right)  }\right]
\exp\left[  ik\tilde{\gamma}\frac{v}{c}\left(  z-\frac{c^{2}}{v}t\right)
\right]  ~,
\end{align*}
where $\psi_{x}$ had been defined by Eq.~(\ref{Xpsi}) and $k$ is a new
parameter---the smallest wavenumber in the spectrum of the pulse. Obviously
$\psi_{fxw}\left(  \rho,z,t\right)  \rightarrow\psi_{x}\left(  \rho
,z,t\right)  $ if $k\rightarrow0$. Due to the second exponential factor,
$\psi_{fxw}$ is not propagation-invariant, while its modulus squared is. \ The energy flow velocity $V_{z}$ along the
$z$-direction evaluated with Eq. (\ref{scalarSw}), that is maximum at $\rho
=0$, does not obey Eq.\ (\ref{R(b)}):  in addition to $v$, it depends on $k$
and $a$, but, interestingly, not on $vt$ arising from the second (subluminal)
speed. For relatively small values of the parameter $a$ and superluminal
values of $v$ close to $c$ Eq.\ (\ref{R(b)}) holds for values of $k$ up to 5
(in reciprocal units of $z,\rho,$ and $a$).

Essentially, the same behavior applies for the energy velocity of the
vector-valued (TM) FXW. In this case however, $V_{z}$ is equal to zero at
$\rho=0$; its maximum value occurs for a value of $\rho>0$. Also, in addition
to $k$, the solution is sensitive to $vt$.

All fields considered so far have infinite total energy, i.e., in reality they
can exist only within a limited aperture. It is interesting to consider the
so-called Modified Focused X wave (MFXW) which has finite energy and is given
by \cite{revPIER}%
\begin{align*}
\psi_{mfxw}\left(  \rho,z,t\right)   &  =\psi_{x}\left(  \rho,z,t\right)
\times\\
&  \left[  \psi_{x}^{-1}\left(  \rho,z,t\right)  +a^{\prime}-i\tilde{\gamma
}\frac{v}{c}\left(  z-\frac{c^{2}}{v}t\right)  \right]  ^{-1}~,
\end{align*}
{\small  }where $a^{\prime}$ is the second width parameter. Again, $V_{z}$,
that is maximum at $\rho=0$, does not obey Eq.\ (\ref{R(b)}) but depends on
$v,a,a^{\prime}$, and $vt$. \ For relatively small values of the parameter
$a$, relatively large values of $a^{\prime}$ and superluminal values of $v$
close to $c$, the formula Eq.\ (\ref{R(b)}) is obeyed very closely because
$vt$ appears as a multiplicative factor of $(v^{2}-c^{2})$.

Essentially, the same behavior applies for the energy velocity of a
vector-valued MFXW. Again, due to the specific construction of the TM field
from the axially oriented Hertz vector potential, in this case $V_{z}$ is
equal to zero at $\rho=0$; its maximum value occurs for a value of $\rho>0$.

\section{Energy flow velocity of subluminal propagation-invariant
electromagnetic and scalar fields}

It is intriguing to ask: if the energy flows always subluminally in a
superluminal pulse, is the flow of the energy of a subluminal pulse faster or
slower than its subluminal group velocity?

The best known and simplest subluminal propagation-invariant scalar field is
the infinite-energy MacKinnon wave packet \cite{MacKinn,revPIER,MeieLorTr}. It is derived from a spherically symmetric standing wave
given by $\exp(-ikct)(\sin kr)/kr$, where $k$ is the wavenumber and
$r=\sqrt{\rho^{2}+z^{2}}$, by applying a Lorentz transformation with
subluminal $\ \beta=v/c$ to the $z$-coordinate and time. For an observer in
another reference frame the field is not any more a monochromatic standing
wave, but a pulse whose intensity distribution propagates with velocity $v$
without any change. We found that the formula Eq.\ (\ref{R(b)}) holds for the
axial energy velocity of the MacKinnon wave packet.

We studied also a finite-energy version of the MacKinnon wave packet given by
\cite{revPIER}%
\[
\psi_{fM}\left(  \rho,z,t\right)  =\frac{\arctan\left[  \frac{\sqrt{\gamma
^{2}\left(  z-vt\right)  ^{2}+\rho^{2}}}{a+i\beta\gamma\left(  z-ct/\beta
\right)  }\right]  }{\sqrt{\gamma^{2}\left(  z-vt\right)  ^{2}+\rho^{2}}}~,
\]
where $\gamma\equiv1/\sqrt{1-(v/c)^{2}}=$ $1/\sqrt{1-\beta^{2}}$ is the common
(subluminal) Lorentz factor and $a$ is a parameter. Again, the formula in Eq.\ (\ref{R(b)}) holds.

As to the vector-valued version of the MacKinnon wave packet, due to the
specific construction of the TM field from the axially oriented Hertz vector
potential with the MacKinnon scalar wavepacket as its z-component, $V_{z}$ is
equal to zero at $\rho=0$; whereas Eq.\ (\ref{R(b)}) holds at its maxima values
that occur at values of $\rho$ corresponding to the maxima of the sinc-function.

The last almost-undistorted spatiotemporally localized field we studied was
the finite energy azimuthally symmetric subluminal splash mode. It arises from
the elementary solution $(\rho^{2}+z^{2}-c^{2}t^{2})^{-1}$ of the scalar wave
equation by first resorting to the complexification $t\rightarrow t+ia$ and
subsequently undertaking a subluminal Lorentz transformation involving the
coordinates $z$ and $t$. A scalar-valued computation yields a maximum axial
energy velocity at the pulse center in conformity with the formula in
Eq.\ (\ref{R(b)}). A vector-valued computation shows that the axial energy
velocity is zero on-axis ($\rho=0$)  at the pulse center. The maximum value of
the axial velocity depends on $vt$ and the parameter $a$. For small values of
$vt$ or subluminal speed $v$ very close to $c$ and $a=1$ the maximum of the
axial energy flow velocity occurs on a ring of radius $\rho=1$.

Finally we studied a propagation-\textit{variant }scalar field, which is
important for applications and is called "pulsed Bessel beam"
\cite{DifRefAxiconBB,PorrGaussjaPBB} Since it is formed by a diffractive
axicon (a circular grating), it is like a disk cut off from a Bessel beam---
its radial profile is propagation invariant and given by the zeroth-order
Bessel function, while its longitudinal profile spreads out in the course of
propagation as a chirped pulse. Such behavior has been experimentally studied
in detail with $%
\operatorname{fs}%
$-range temporal and $%
\operatorname{\mu m}%
$-range spatial resolution \cite{meieDifAxicon0,meieDifAxicon1}.

In the given case, the field to be inserted into Eq.~(\ref{scalarSw})
factorizes as $\psi=U(z,t)J_{0}(k_{0}\rho\sin\theta)$, where $k_{0}$ is the
carrier wavenumber (mean wavenumber in the spectrum of the pulse). For the
field of a pulsed Bessel beam with Gaussian temporal profile, the function
$U(z,t)$ was calculated using Eqs. (44), (45) and (54) from
Ref.~\cite{PorrGaussjaPBB}. As one can see in Fig.~4, the pulse broadens and
gets chirped in the course of propagation. The reason is that waves of
different wavenumbers diffract at different angles on the grating, which means
that the Axicon angle is not constant over the spectrum of the pulse.

\begin{figure}
\centering
\includegraphics[width=8.5cm]{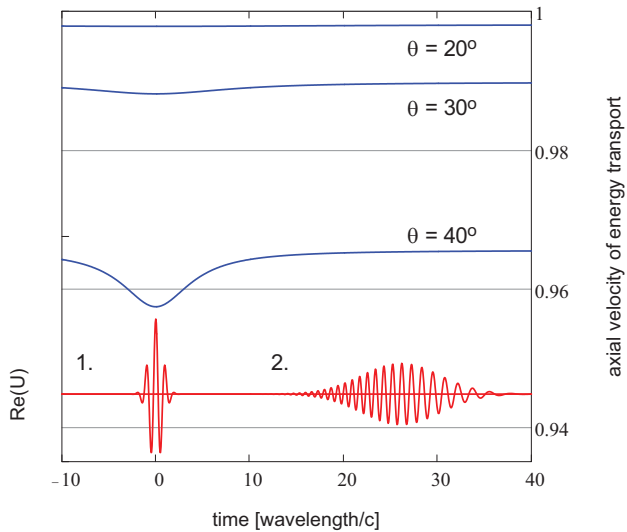}%
\caption{Time dependence of the real part of
the function $U(z,t)$ of a pulsed Bessel beam with a mean value of the Axicon
angle $\theta=40^{\circ}$ (determined for the carrier wavenumber) and Gaussian
temporal profile of width $1$ (in units of carrier wavelength divided by $c$):
curve 1 -- at the origin $z=0$ where the pulse is formed and has the shortest
duration, curve 2 -- at $z=20$ (in units of the carrier wavelength). The
curves with indicated mean Axicon angles show temporal behavior of the energy
velocity (normalized to $c)$ at the maximum of the pulse envelope.}
\end{figure}

Despite the chirp, locally the pulse looks like a Bessel beam characterized by
an instantaneous wavenumber and corresponding Axicon angle. Therefore, based
upon the results obtained above, Eq.~(\ref{R()}) should hold at time instances
and propagation distances when the pulse contains more than just a few cycles.
This is exactly what we see in Fig.~4: at small mean angles $\theta$---not
speaking about paraxial angles---the energy velocity is almost at a constant
level determined by Eq.~(\ref{R()}) \ The slight drop in its value occurs only
at the origin if the pulse is shorter than 2 wavelengths there and
$\theta>20^{\circ}$. The drop at such extreme parameters may be caused also by
the circumstance that in the calculation of the function $U(z,t)$ in Ref.
\cite{PorrGaussjaPBB} the group velocity dispersion is only approximately taken into account.

To conclude, the answer to the question raised in the beginning of the Section
is: in those regions of a subluminal pulse where Eq.~(\ref{R(b)}) holds, the
energy flows faster than the pulse envelope, i.e. $0<v<V_{z}=V<c$

\section{Discussion}

We have seen that---with a few exceptions---the formula given in
Eq.~(\ref{R(b)}) holds for the energy axial velocity of both superluminal and
subluminal non-diffracting wavefields. Moreover, Eq.~(\ref{R(b)}) indicates
that the velocity does not change if one makes a transition
\textit{superluminal}$\longleftrightarrow$\textit{subluminal}, i.e., replaces
the normalized group velocity $\beta=v/c$ by its reciprocal value $c/v$. To
understand the reason of such an interesting feature of the expression let us
take a look at the expression of the energy velocity in terms of impedance
\cite{ImpedZ}, \textit{viz}.%
\begin{align}
\mathbf{V}& =2c\frac{\mathbf{E}\times c\mathbf{B}}{\mathbf{E}^{2}+c^{2}%
\mathbf{B}^{2}}=c\frac{2Z}{1+Z^{2}}\left( \mathbf{e}_{E}\times \mathbf{e}%
_{B}\right) ~,  \label{VZ} \\
Z& \equiv \frac{E}{cB}=\frac{E}{H}Z_{0}^{-1}~,  \label{Zdef}
\end{align}%
where $E$, $B$ are the magnitudes of electric and magnetic field vectors and
$\mathbf{e}_{E}$, $\mathbf{e}_{B}$ are corresponding unit vectors; $Z$ is the
impedance normalized to the vacuum impedance $Z_{0}=\sqrt{\mu_{0}%
/\varepsilon_{0}}\approx377\Omega$. Thus, the group velocity is determined by
the impedance $\beta=Z$ and the energy velocity Eq.~(\ref{VZ}) does not change
if we insert $Z^{-1}$ instead of $Z$. This is due to the invariance of the
energy flux and energy density with respect to the duality transformation
$\mathbf{E}\rightarrow c\mathbf{B}$, $\mathbf{B}\rightarrow-\mathbf{E/}c$ .
From the duality also follows that our results obtained for TM pulses apply
also for TE pulses. $Z=1$ and, consequently, $V=c$ holds only for TEM and
\textit{null }electromagnetic waves \cite{Lekner2002,YannisLWII,Kaiser2011}.

Note also that Eq.~(\ref{R(b)}) resembles the relativistic composition law for
velocities. One can speculate that the energy velocity equals the pulse
propagation velocity seen from a reference frame countermoving with exactly
the pulse group velocity. 

It is well known that the Poynting vector is not defined uniquely by the
Poynting theorem. Could it be that, consequently, the energy flow velocity we
have dealt with throughout this paper is not also defined uniquely and
therefore its nonequality to the group velocity would not be of interest? The
answer is that any other definition of the Poynting vector might violate the
velocity upper limit $c$ \cite{Lekner2002}. Here a quotation from
\cite{Jackson}, section 8.5 is appropriate: "However the theory of special
relativity, in which energy and momentum are defined locally and invariantly
via the stress--energy tensor, shows that the ... expression for the Poynting
vector is unique."

According to conventional thinking, energy should be tightly coupled to an EM
field pulse. How, then, one has to interpret the results that energy flows
slower than a superluminal pulse itself and flows faster than a subluminal
pulse? Nonequality of energy flow velocity to the velocity of field motion is
not unique for non-diffracting localized waves but takes place in other
non-null fields, e.g., in standing waves, dipole radiation, \cite{Kaiser2011},
etc. In Ref.~\cite{Kaiser2011} this nonequality is explained in terms of
reactive (rest) energy that the field leaves behind. In this paper one can
also find a hint towards comprehension of this inequality: "A rough way to
understand why $V<c$ is by analogy with water waves. The mass carried by the
waves has a definite speed at each point and time, but this need not coincide
with the propagation speed of the wavefronts." Existence of a rest energy
portion in non-diffracting localized waves is obvious because all of them
contain a standing-wave component. It is interesting to note that the standing
wave component inherent to these waves can be given an interpretation as if
the mass of a photon of these wavefields is not equal to zero
\cite{Minupeatykk}.

Finally, let us note that the signal velocity of superluminal nondiffracting
pulses is not superluminal but an instantaneous notch made, e.g., into
Bessel-X pulse, which transforms into the subluminal propagation-variant pulse
considered above, as proved by a thought experiment in Ref.~\cite{Minupeatykk}.

\section{Conclusions}

We have shown that the velocity $V$, with which energy in non-diffracting
pulsed waves flows in the direction of propagation, is not equal to the
propagation velocity $v$ (group velocity) of the pulse itself. Instead, on the
symmetry axis and/or at the locations of the energy density maxima, these two
quantities obey a simple but physically content-rich relation $V=2v/\left[
1+(v/c)^{2}\right]  $. This has been proven first for vector-valued
superluminal 2D light sheets and their 3D cylindrical generalizations in Sec. III. Subsequently, it has been shown to be valid for the scalar-valued $m$-order
Bessel beam, the fundamental zero-order X wave, the first-order azimuthally
asymmetric X wave, as well as the corresponding vector-valued TM
electromagnetic fields based on a Hertzian potential approach. The behavior of
the axial velocity for the vector-valued fields differs depending on whether
the scalar potential used as a seed in forming the vector Hertz potential is
azimuthally symmetric or asymmetric. A detailed discussion is provided in Sec. IV.

Purely propagation-invariant fields characterized by the group speed $v$ along
the direction of propagation are physically unrealizable. Physically
realizable spatiotemporally localized waves contain two speeds: the group
speed $v$ (superluminal or subluminal) and a second speed $c^{2}/v$
(subluminal or superluminal). Between the purely propagation-invariant and the
physically realizable \textquotedblleft almost undistorted\textquotedblright%
\ localized waves, there exists a family of only intensity-invariant localized
waves containing the aforementioned two speeds. Examples for such pulses are
the Focus X Wave (FXW) and MacKinnon's wave packet. Both are characterized by
infinite energy content. The energy flow velocity for these two scalar fields,
the scalar finite-energy pulses based on them, as well as the corresponding
vector-valued TM electromagnetic fields determined by a Hertzian potential
approach obey the universal formula given in Eq. (\ref{R(b)}) very closely
provided the group velocity is very close to the speed of light and certain
free parameters are tweaked appropriately. Specific details are given in Secs. III-V.

It is very interesting to note that the universal formula for the axial energy
flow velocity, in the form appearing in Eq. (\ref{R(b)}), is intimately
related to the wave impedance reformulation in Eq. (\ref{VZ}), which, in turn,
is reminiscent of a relativistic expression for the addition of velocities.

Finally, a note on superluminality is appropriate. The presence of a
superluminal speed in a finite-energy solution does not contradict
relativity. If the parameters are chosen appropriately, the pulse moves
superluminally with almost no distortion up to a certain distance $z_{d}$,
which is determined by the geometry (aperture size and an axicon angle) and
then it slows down to a luminal speed $c$, with significant accompanying
distortion. Although the peak of the pulse does move superluminally up to
$z_{d}$, it is not causally related at two distinct ranges $z_{1},z_{2}%
\in\lbrack0,z_{d})$. Thus, no information can be transferred superluminally
from $z_{1}$ to $z_{2}$. The physical significance of such wavepackets is due
to their spatiotemporal localization.

The authors thank Ari Friberg and John Lekner for giving comments to their
papers \cite{Ari1993} and \cite{Lekner2002}, which stimulated undertaking the
present study.

\appendix

\section{PROOF OF CONDITION $\left\vert \mathbf{V}\right\vert \leq c$ FOR
SCALAR FIELDS}

Similarly to Eq.~(\ref{v<c}) we can write by using Eq.~(\ref{scalarSw}):%

\begin{equation}
1-\frac{\mathbf{V}^{2}}{c^{2}}=1-\frac{1}{c^{2}}\frac{\left(  \dot{\psi}%
^{\ast}\mathbf{\nabla}\psi+\dot{\psi}\mathbf{\nabla}\psi^{\ast}\right)  ^{2}%
}{\left(  \frac{1}{c^{2}}\dot{\psi}\dot{\psi}^{\ast}+\mathbf{\nabla}\psi
\cdot\mathbf{\nabla}\psi^{\ast}\right)  ^{2}} \label{A1}%
\end{equation}

This form can be used to prove that $\left\vert \mathbf{V}\right\vert \leq c$.
\ Indeed, making use of the notations $F\equiv\mathbf{\nabla}\psi
\cdot\mathbf{\nabla}\psi^{\ast}-\frac{1}{c^{2}}\dot{\psi}\dot{\psi}^{\ast}$
and $\mathbf{G}\equiv\dot{\psi}\mathbf{\nabla}\psi^{\ast}$ for convenience, we
can transform Eq.~(\ref{A1}) as follows:%
\begin{align*}
1-\frac{\mathbf{V}^{2}}{c^{2}} &  =\frac{F^{2}+\frac{2}{c^{2}}\mathbf{G}%
\cdot\mathbf{G}^{\ast}-\frac{1}{c^{2}}\left(  \mathbf{G}\cdot\mathbf{G}%
+\mathbf{G}^{\ast}\cdot\mathbf{G}^{\ast}\right)  }{\left(  \frac{1}{c^{2}}%
\dot{\psi}\dot{\psi}^{\ast}+\mathbf{\nabla}\psi\cdot\mathbf{\nabla}\psi^{\ast
}\right)  ^{2}}=\\
&  =\frac{F^{2}+\frac{2}{c^{2}}\mathbf{G}\cdot\mathbf{G}^{\ast}+\frac{4}%
{c^{2}}\left(  \operatorname{Im}\mathbf{G}\right)  ^{2}-\frac{2}{c^{2}%
}\mathbf{G}\cdot\mathbf{G}^{\ast}}{\left(  \frac{1}{c^{2}}\dot{\psi}\dot{\psi
}^{\ast}+\mathbf{\nabla}\psi\cdot\mathbf{\nabla}\psi^{\ast}\right)  ^{2}}\\
&  =\frac{F^{2}+\frac{4}{c^{2}}\left(  \operatorname{Im}\mathbf{G}\right)
^{2}}{\left(  \frac{1}{c^{2}}\dot{\psi}\dot{\psi}^{\ast}+\mathbf{\nabla}%
\psi\cdot\mathbf{\nabla}\psi^{\ast}\right)  ^{2}}~.
\end{align*}
The denominator and numerator of the last fraction are nonnegative and, as a consequence, $\left\vert \mathbf{V}\right\vert \leq c$.

\end{document}